\documentclass[twocolumn]{aastex62}
\pdfoutput=1
\usepackage{amsfonts,amsmath,amssymb}
\usepackage{graphicx}
\usepackage{color}
\usepackage{tabularx}
\usepackage{xcolor}
\usepackage{natbib}
\usepackage[bookmarks=false]{hyperref}
\usepackage{enumitem}
\defcitealias{Shields}{S86}

\begin{document}

\title{On the wind-driven relaxation cycle in accretion disks}

\author{Shalini \surname{Ganguly}} \altaffiliation{E-mail: ganguly@unlv.nevada.edu}
\author{Daniel \surname{Proga}} \altaffiliation{E-mail: dproga@physics.unlv.edu}

\affiliation{Department of Physics and Astronomy, University of Nevada Las Vegas}

\begin{abstract}
A disk wind can cause perturbations that propagate throughout the disk
via diffusive processes. On reaching the inner disk, these perturbations can change
the disk luminosity, which in turn, can change the wind mass loss rate, $\dot{M}_w$.
It has been argued that this so-called ``wind driven relaxation cycle"  might explain
the observed variability in some disk accreting objects.
Here, we study the response of the innermost mass accretion rate $\dot{M}_a$ to the loss of matter at different rates and radii. We allow the wind launching radius, $R_L$,
to scale with $\dot{M}_a$. We computed a grid of time-dependent models for
various $\dot{M}_w$-$\dot{M}_a$ and $R_{L}$-$\dot{M}_a$ dependencies.
We find that the disk behaviour significantly differs
for the `variable $R_L$' case  compared to the `fixed $R_L$' case.
In particular, much stronger winds are required to destabilize the disk
in the former than the latter case.
However, the $\dot{M}_a$ amplitude does not grow significantly
even for unstable cases because the oscillations saturate at a low level
either due to disk depletion or due to the wind being launched at
very small radii, or both. This result implies that disk winds
are unlikely to be responsible for state transitions as those require large
changes in the inner disk. Despite modest changes at the inner disk regions, the disk surface density at large radii can vary with a large amplitude, i.e.,
from 0 to a few factors of the steady state value. 
This dramatic variation of the outer disk could have observable consequences.\\
\vspace{1cm}
\end{abstract}

\section{Introduction}
\label{sec:1}

Various astrophysical systems including X-ray binaries, young stellar objects (YSOs), cataclysmic variables (CVs) and active galactic nuclei (AGNs) are powered by accretion disk processes. The relatively high luminosity of these objects is due to the efficient conversion of accretion power into radiation. The luminosity generated from such accretion processes tend to vary with time which provides us with important clues as to the nature of the accretion disk, the accretor, and also the object or source supplying matter to the disk.

For a constant rate of mass supply, time variability could be attributed to spatial and temporal variations in the disk structure  or the strength and configuration of the disk magnetic field. Such variations can arise due to the extended nature of the disks, with the inner and outer radii differing by orders of magnitude (ranging between two orders in CVs, to about seven orders in AGNs). The surface properties at small radii differ from those at large radii and there is a host of different physical processes that may cause the accreting material to undergo time-dependent evolution. They include a variety of local instabilities such as convective, thermal or magneto-rotational instability \citep{BH98,FL19}. Yet there are several non-local processes that could also lead to time variability.

One of the consequences of the large radial extent is the huge variation in escape velocity throughout the disk. Thus, as long as a disk has a slightly concave surface, the  high-energy radiation that is emitted by the inner disk can irradiate the outer disk leading to the formation of a high-temperature surface layer in which thermal speeds can exceed the escape velocity. This can drive a strong wind from outer radii and cause a disruption in the accretion flow. When information about this disruption reaches the inner disk, it changes the local emission which, in turn, affects the disk irradiation. Thus, the radiation from the inner disk acts as a coupling between the inner and outer disks. The self-irradiated disk is an example of a ``self-regulated accretion" process \citep[][S86 hereafter]{Shields} with feedback.

The Compton-heated corona and eventual disk wind \citep{Shields1} have widespread applications in understanding the absorption lines observed in AGNs \citep{W96} and  X-ray binaries \citep{PK02,2010Lu,M15,WP18}. However, it is unclear as to what degree a disk wind can destabilize the accretion disk and be responsible for the observed variability in the luminosity and spectral energy distribution (SED). To assess this role of the disk winds, we may define a variable $\eta_{ w} \equiv \dot{M}_{ w}/\dot{M}_{ a}$, where $\dot{M}_{ w}$ and $\dot{M}_{ a}$ are the wind mass loss rate (at the outer disk region) and mass accretion rate onto an accretor, respectively. The ratio measures the efficiency of wind driving due to the accretion power and indicates how strongly the wind is coupled to the latter. The model of instantaneous response of wind-to-accretion and vice versa showed that a wind with $\eta_{ w}$ as low as one, destabilizes the disk \citep{Shields1}. However, taking the effect of viscosity into account, \citetalias{Shields} found that the accretion at the inner disk edge responded much slower to the change in the disk surface density, $\Sigma$, at large radii. Viscosity stabilizes the disk by producing a ``delay" or ``relaxation time" to the propagation of perturbations throughout the disk. Therefore, a much higher $\eta_{ w}$ was needed to generate variability in the disk.

For systems with relatively high luminosities, the escape velocity from a disk at a given radius could be reduced by the radiation pressure on free electrons and due to opacity from spectral lines and bound-free processes. In X-ray binaries, the latter two are negligible because the gas is highly ionized and few lines as well as few bound transitions are present. Yet, as expected, and shown both numerically and theoretically, the radiation pressure on free electrons introduces a linear scaling between the launching radius of the thermal wind and the luminosity, $L$  \citep{PK02}. In AGNs and CVs, the radiation pressure on lines (line driving) can produce a wind with $\dot{M}_{ w} \propto L^{1/\alpha}$, where $\alpha$ is the force multiplier parameter. The value of $\alpha$ depends on the SED, the chemical composition, and the physical conditions in the gas but it generally ranges from ~0.2 to 0.8 \citep[e.g.,][]{CAK}. The above scaling is quite universal as it holds for 1-D stellar wind \citep{CAK} as well as for 2-D \citep[e.g.,][]{PSD98, P99} and even 3-D disk winds \citep{DP18}.

In this paper, we study a more generalized model of disk wind coupling to accretion power. We verify our results against the classic case of linear dependence of wind on accretion rate \citepalias{Shields} and then extend our analysis to non-linear dependencies of the wind. This allows us to test our model against different possibilities. Our main focus is to study the dynamic variability of the launching radius of the wind and explore our model for different free parameters. The outline of the paper is as follows: In section \ref{sec:2}, we describe in brief the mathematical and computational techniques used in our analysis. In section \ref{sec:3}, we introduce our calculation and verification of the \citetalias{Shields} result (with increased numerical resolutions). Our generalized approach towards the problem allows us to look at the results of two different models holistically and identify several previously unexplored cases. Finally, in section \ref{sec:4}, we summarize the applications of disk oscillations that have been studied in the past and what future prospects it might hold.

\begin{figure}
  \centering
  \includegraphics[scale=0.5]{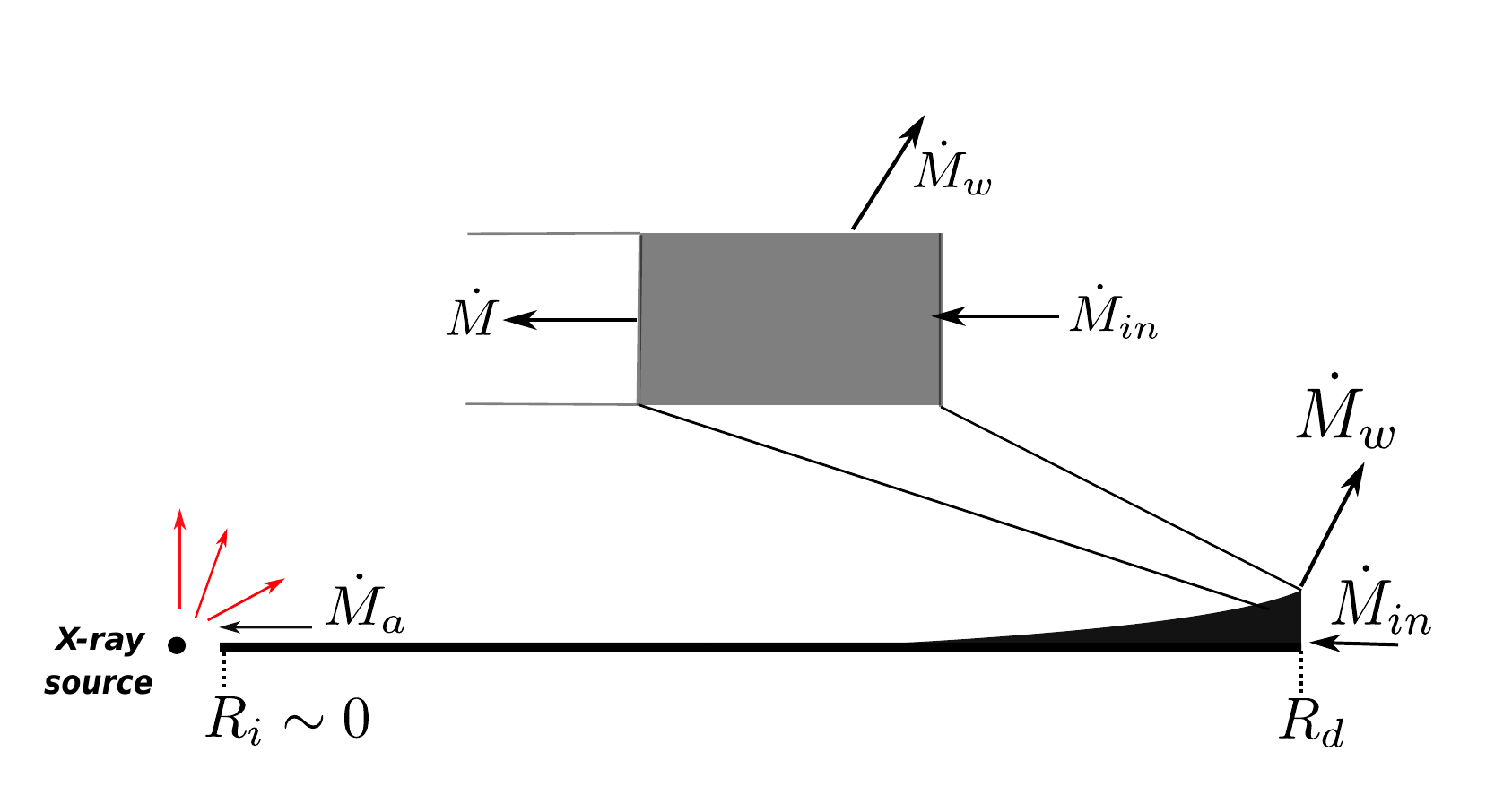}
  \caption{Schematic of the model described as ``\textit{$\delta$-function wind at $R_d$ with $\nu\propto R$}" studied in S86. Matter enters the disk at a constant rate at $R_d$ while the X-ray source accretes matter from the innermost disk radius. The accretion leads to X-ray irradiation of outer disk region which drives the wind. The inset shows the outermost radial grid, where diffusion of matter at a rate $\dot{M}$ and removal of matter by a wind at rate $\dot{M}_w$ are compensated by mass injection ($\dot{M}_{in}$). Diffusive process disperses matter throughout the disk.}
  \label{fig:scheme}
\end{figure}

\section{Methods}
\label{sec:2}

\subsection{Equations and analytical results}
\label{sec:2.a}

We assume azimuthal symmetry and perform 1D simulations on a geometrically thin and optically thick disk along the radial direction. We have adopted similar formulations and notations as used in \citetalias{Shields}. Their approach involves solving the diffusion equation that describes the disk evolution \citep{P81,1974L}. As in S86, we assume a constant rate of mass injection at the outermost disk radius, $R_d$, from an external source. This is true for all the cases studied henceforth. The loss of mass in the form of wind takes place at the very same radius for the first model, which corresponds to the model described by eq.(3.5) in S86 (see $\S$III.(b) in S86). Fig.~\ref{fig:scheme} shows a schematic of this model. We also assume a simple radius-dependent viscosity $\nu \propto R$, similar to that used in \cite{1974L}, for all our models. The mass continuity and angular momentum conservation gives us the following diffusion equation,
\begin{align}
    &\frac{\partial \Sigma}{\partial t} = \frac{1}{2\pi R} \frac{\partial \dot{M}}{\partial R} + S(R,t) \textrm{,}\label{eq4}
\end{align}
where,
\begin{align}
    \dot{M} =& 6\pi R^{1/2} \frac{\partial}{\partial R}(\nu \Sigma R^{1/2}) \label{eq5} \qquad \textrm{and,}\\
    S(R,t) =& S_{in}(R) - S_w(R,t)\textrm{.}
\end{align}
Here $\dot{M}$ is the mass accretion rate at a given radius, $\nu$ is the kinematic viscosity and $\Sigma$ is the local surface mass density. The net source term, $S$, accounts for both a steady mass input to the disk ($S_{in}$), as well as a mass loss due to wind ejection from the disk ($S_w$). The general mass input and output rates respectively, are defined by
\begin{align}
    \dot{M}_{in} = 2\pi \int_{R_i}^{R_d} S_{in}(R) R\quad dR \qquad \textrm{and,}
\end{align}
\begin{align}
    \dot{M}_w(<R) = 2\pi \int_{R_i}^{R} S_w(R^\prime) R^\prime\quad dR^\prime \textrm{,}
\end{align}
where $R_i$ and $R_d$ are the innermost and outermost disk radii, respectively, and the mass loss is calculated up to an arbitrary radius R within which mass is being lost. The above equations are recast using dimensionless variables as in eqs. (2.5)-(2.10) in S86. We present these equations below for clarity of our method description,
\begin{align}
    R_* \equiv & \quad R/R_d\textrm{,}\\
    \nu_* \equiv & \quad \nu/\nu_0\textrm{,}\\
    \dot{M}_* \equiv & \quad \dot{M}/\dot{M}_{in}\textrm{,}\\
    \Sigma_* \equiv & \quad \Sigma/\Sigma_0\textrm{,} \\
    t_* \equiv & \quad t/t_0\textrm{,}\\
    S_* \equiv & \quad SR_d^2/\dot{M}_{in}\textrm{,}
\end{align}
where $\nu_0$ is the characteristic viscosity, $\Sigma_0 \equiv \dot{M}_{in}/\nu_0$ is the characteristic surface density and $t_0 \equiv R_d^2/\nu_0$ is the characteristic viscous time scale. Using the coordinate transformation, $x = R_*^{1/2}$, equations \ref{eq4} and \ref{eq5} can be rewritten as,
\begin{align}
    \frac{\partial \Sigma_*}{\partial t_*} =& \frac{1}{4\pi x^3} \frac{\partial \dot{M}_*}{\partial x} + S_*(x,t) \textrm{,} \label{eq:diff}\\
    \dot{M}_* =& 3\pi \frac{\partial}{\partial x}(x\nu_* \Sigma_*) \textrm{.}
\end{align}
As in the wind model studied by \citetalias{Shields}, we treat mass loss using a delta function (launching of wind at a given radius), and \citetalias{Shields} defined the main model parameter as 
\begin{align}
    C \equiv& \dot{M}_w/\dot{M}_a \textrm{.}\label{simrel}
\end{align}
where $\dot{M}_a \equiv \dot{M}(R_{in})$. 
In this paper, we alternatively refer to this ratio as the wind efficiency, $\eta_w$. While discussing or referring to the classic case studied in \citetalias{Shields}, we use the `$C$' notation for comparison. As demonstrated by \citetalias{Shields}, once the disk has attained a steady state, perturbations in disk surface density would either persist, grow, or decay with time. This forms the basis of the ``self-regulated accretion" and $\eta_w$ (or $C$) determines how strongly coupled the wind is to the central accretion rate and hence, to the luminosity of the central source. At steady state, the mass conservation relation applied to the disk, gives
\begin{align}
    \dot{M}_{in} =& \dot{M}_a^{(s)} + \dot{M}_w^{(s)} \textrm{,}
\end{align}
where the superscript $(s)$ denotes the steady state value. The expression for steady state mass accretion rate normalized to the mass input rate $\dot{M}_{in}$ reads in the following way,
\begin{align}
    \dot{M}_{a*}^{(s)} =& \frac{1}{1+C} \textrm{.}
\end{align}
Using a straightforward radius-dependent viscosity law, \citetalias{Shields} performed an analytical calculation to find the critical value of $C$ that would lead to a perpetual oscillation in disk density about its steady state value. The amount of matter depleted through wind and accretion is continually replenished by the constant supply of matter. This results in a variability of mass accretion rate and luminosity at the inner edge of the disk. For the parameter $C$, \citetalias{Shields} obtained the value required for sustained stable (critical) oscillations analytically as, 
\begin{align}
    C_{crit} = \cosh \pi \approx 11.6 \label{eq3} \textrm{,}
\end{align}
such that when $C < C_{crit}$, the oscillations decay, whereas for $C > C_{crit}$, they grow.

The wind mass loss rate might not be a linear function of mass accretion rate. \citetalias{Shields} considers such a possibility and they show based on their analytical treatment that for a power-law dependence, $C \propto \dot{M}_a^{k_l}$, where $k_l$ is an arbitrary constant, the $C_{crit}$ would be reduced by a factor of $(1+k_l)^{-1}$ (see analogous equation 4.5 in \citetalias{Shields}). The case with $k_l=0$ corresponds to the case discussed above, where equation \ref{eq3} gives the value of $C_{crit}$. 

We formally approach this possibility by generalizing our equation \ref{simrel} and writing it in the following way:
\begin{align}
    \dot{M}_{w*} \equiv C^\prime \dot{M}_{a*}^p \textrm{.}\label{eq7}
\end{align}
where the subscript $*$ stands for mass loss rates normalized to the input mass rate $\dot{M}_{in}$, where any remaining constant of proportionality has been absorbed into $C^\prime$, and $p$ is an arbitrary constant exponent. Note that equation \ref{eq7} reduces to equation \ref{simrel} when $p=1$. The requirement for the steady state condition becomes
\begin{align}
    \dot{M}_{a*}^{(s)} +& C^\prime (\dot{M}_{a*}^{(s)})^p - 1 = 0 \textrm{,}\label{eq1}
\end{align}
but this equation needs to be solved numerically.

In the above analyses, a fixed wind launching radius has been assumed. However, in general, this may not be an ideal condition. Our main focus here is to examine the effects of relaxing this assumption by allowing the radiation from the inner disk to irradiate the entire disk and causing reduced local escape velocity. We can express luminosity $L$ in units of the Eddington factor $\Gamma$, such that $L = \Gamma L_{Edd}$, where $L_{Edd}$ is the Eddington luminosity. Here we assume that the irradiation luminosity $L$ of the disk equals to the total accretion luminosity. We express the coupling between the launching radius and $\dot{M}_{a}$ using the following expression:
\begin{align}
    R_{L*} = 1 - \Gamma \frac{\dot{M}_{a*}}{\dot{M}_{a*}^{(s)}} \textrm{,}\label{eq2}
\end{align}
where $R_{L*}$ is the launching radius normalised to $R_d$. Our equation \ref{eq2} is similar to equation (22) in \citet{PK02} that was derived for the launching radius of a Compton-heated wind, corrected for radiation driving. When $\Gamma=0$, $R_L=R_d$ and we have the case studied by \citetalias{Shields}. The only numerical constraint is that $R_L\geq0$.

Using their analytical method for the simplest case ($\Gamma=0$, $p=1$), \citetalias{Shields} derived an expression for the period of oscillations as
\begin{align}
    P_* \propto \frac{R_L}{R_d}
\end{align}
We expect that in our variable $R_L$ model, the disk stability condition and the variability period will be sensitive to $\Gamma$.

\subsection{Numerical methods}
\label{sec:2.b}

We have used numerical methods similar to that used in \cite{BP}. We have developed a Python code to study the effects of self-regulated accretion as discussed in the previous section. The wind launching zone is a delta function, with mass loss taking place from a single radial grid zone.
\begin{align}
    S_{w*} = \frac{C\dot{M}_{a*}}{\pi (R_{j*}^{\prime 2}-R_{j-1*}^{\prime 2})} \label{eq:wind}
\end{align}
where $R_{j*}^\prime=x_j^{\prime 2}$ and $x_j^\prime = (x_j + x_{j+1})/2$. Here $R_j$ refers to the wind launching radius and $R_{j*}^\prime$ is the averaged launching radius value used to calculate $S_w$. The disk surface density $\Sigma_*$ is assigned an initial value of 0. We take $\dot{M}_{in}=1$, since all mass rates are normalised to $\dot{M}_{in}$ and together with eq.\ref{eq:wind}, calculate the source term $S_*$. The diffusion equation \ref{eq:diff} is then solved using forward difference method while updating $\Sigma_*$. The boundary condition is obtained by imposing the condition that mass flux is finite and conserved at the disk edges, i.e. $\partial(\nu\Sigma)/\partial R = 0$.

Our resolution study shows that the calculated value of the accretion rate depends on the width of each radial zone. In particular, the variable $\Gamma$ model is sensitive to smaller resolution in $x$, $N_x$ ($N_x<$ 200). Hence, for most of our models presented below, we use $N_x=200$ (see $\S$\ref{sec:3} for more discussion). 

To ensure that the value of $\Sigma$ remains realistic at all times, it is quite common to impose the condition $\Sigma_*=0$, whenever the numerical solution leads to $\Sigma_*<0$. This condition is especially important when we deal with growing oscillations. Instead of allowing the disk to deplete completely, we introduce a floor value of 0.001 for $\Sigma$.

Three time scales that enter this problem are: 1) The mass outflow time scale, $t_{\textrm{out}}$, 2) mass inflow time scale, $t_{\textrm{in}}$ and, 3) diffusion time scales, $t_{\textrm{diff}}$. These scales can be defined as
\begin{align}
    t_{\textrm{out}} = \dot{\Sigma}(R_L)/\dot{\Sigma}_w(R_L)\textrm{,}\\
    t_{\textrm{in}} = \dot{\Sigma}(R_{\textrm{in}})/\dot{\Sigma}_{\textrm{in}}(R_L)\textrm{ and,}\\
    t_{\textrm{diff}} = 0.25 \frac{4\Delta x^2}{3}\label{eq:t} \textrm{.}
\end{align}
where eq. \ref{eq:t} expresses  the stability criterion of eq. \ref{eq:diff} which is a diffusion equation (see e.g., Press et al. 2007). In our numerical calculations, we choose a time step to be 20$\%$ of the shortest of the above three time scales.

The initial condition is to set the disk surface density to be 0 and let matter diffuse from the surrounding source until the accretion disk reaches a steady state. We then perturb it by switching on the wind. In practice, we allow the system to reach $(1-\epsilon)\dot{M}_{a*}^{(s)}$, where $\epsilon$ is a very small number of the order of $10^{-3}$. It takes roughly 3 viscous time scales to reach this near steady state. We use the \textit{fsolve} method in SciPy's \footnote{Python 3.6.7} optimization library to find the root of equation \ref{eq1}.

\section{Results}
\label{sec:3}

\begin{figure}[ht]
\includegraphics[scale=0.5]{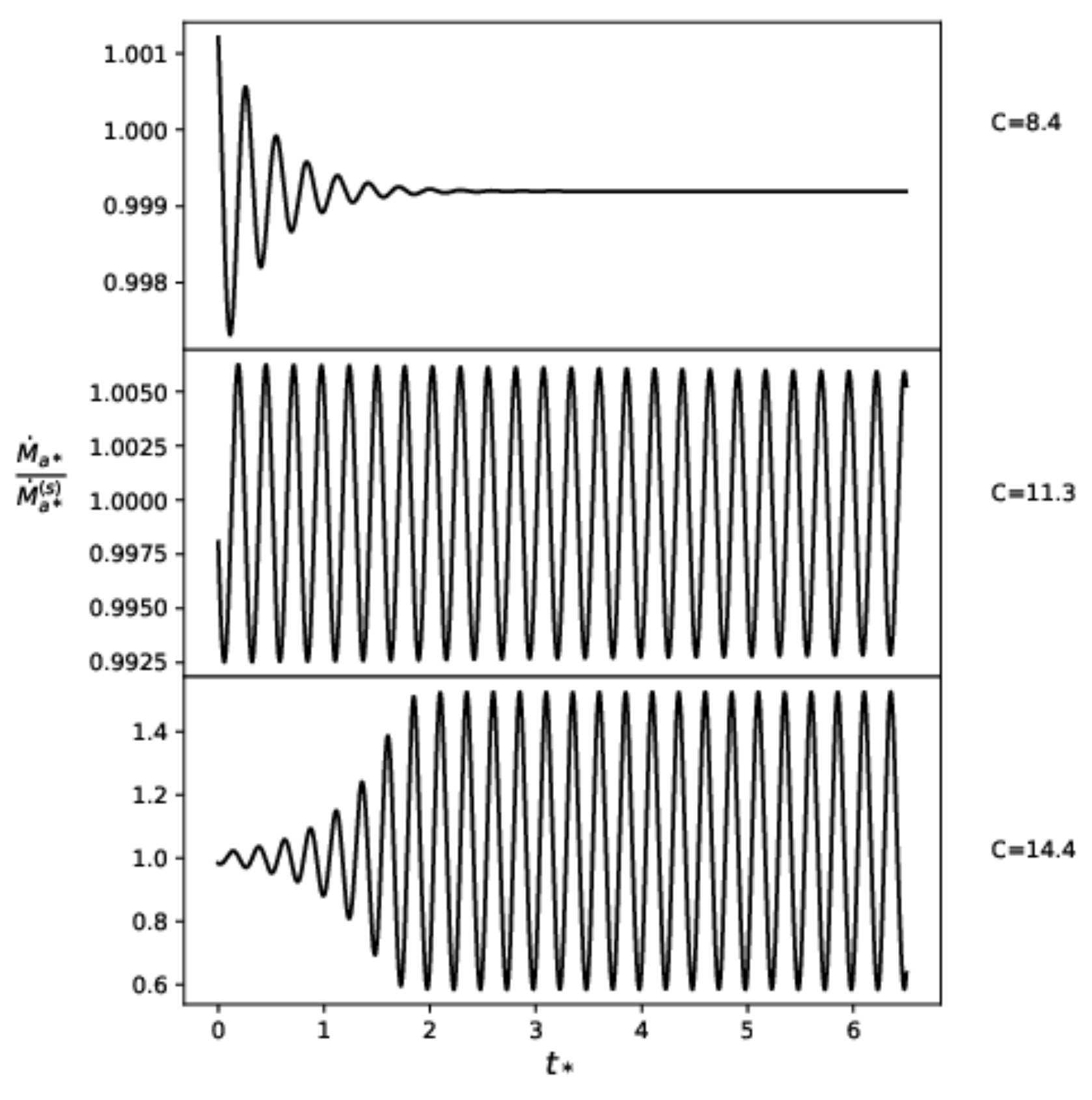}
\caption{Central mass accretion rate (normalized to the steady state mass accretion rate) evolution of the disk for different wind strength parameter $C$. The top panel shows decayed oscillations in the disk, the middle panel shows the critical $C$ case where the oscillation persists with constant amplitude. The bottom panel shows a growing oscillation phase, which saturates after some time due to local disk depletion.}
\label{fig:shields result}
\end{figure}

First we checked the results from our simulations against the results presented in \citetalias{Shields}. Our resolution is higher than theirs by a factor of 10. This increased resolution does not considerably affect the key result although it does substantially reduce the amplitude of oscillation. This can be attributed to the fact that we have limited our analysis to a single zone wind regardless of the radial resolution. Namely, we still inject and eject the same amount of matter but now from a radial ring with smaller area. We find that $C_{crit} \approx 11.3$ as opposed to 11.4 obtained by \citetalias{Shields}. Fig. ~\ref{fig:shields result} illustrates our results in a similar manner to the fig.$\sim$ 1 by \citetalias{Shields}. As expected, we observe damped oscillation for $C<11.3$ and growing oscillations for $C>11.3$. We note that we consider only those oscillations to be stable whose amplitudes vary by less than $10^{-4}$. This condition is consistently followed throughout our analysis, and is used to classify the oscillations into categories. 

In the growing oscillation cases, the mass accretion rate grows until it saturates after some time. This saturation of oscillation is caused by the local surface density reaching negligible values or, in other words, complete depletion of matter from that region of the disk. We expect that if S86 continued their calculations to longer times, they would likely find the same behaviour. For $C=14.4$ shown above, our simulations showed that during this phase, $\Sigma_{*,min}$ approaches 0, whereas $\Sigma_{*,max}$ approaches 2.5 times its steady state value. The outcome of this instability is a small amplitude $\dot{M}_{a*}$ oscillation but a large modification of the disk solution. 

\begin{figure}[ht]
\includegraphics[scale=0.55]{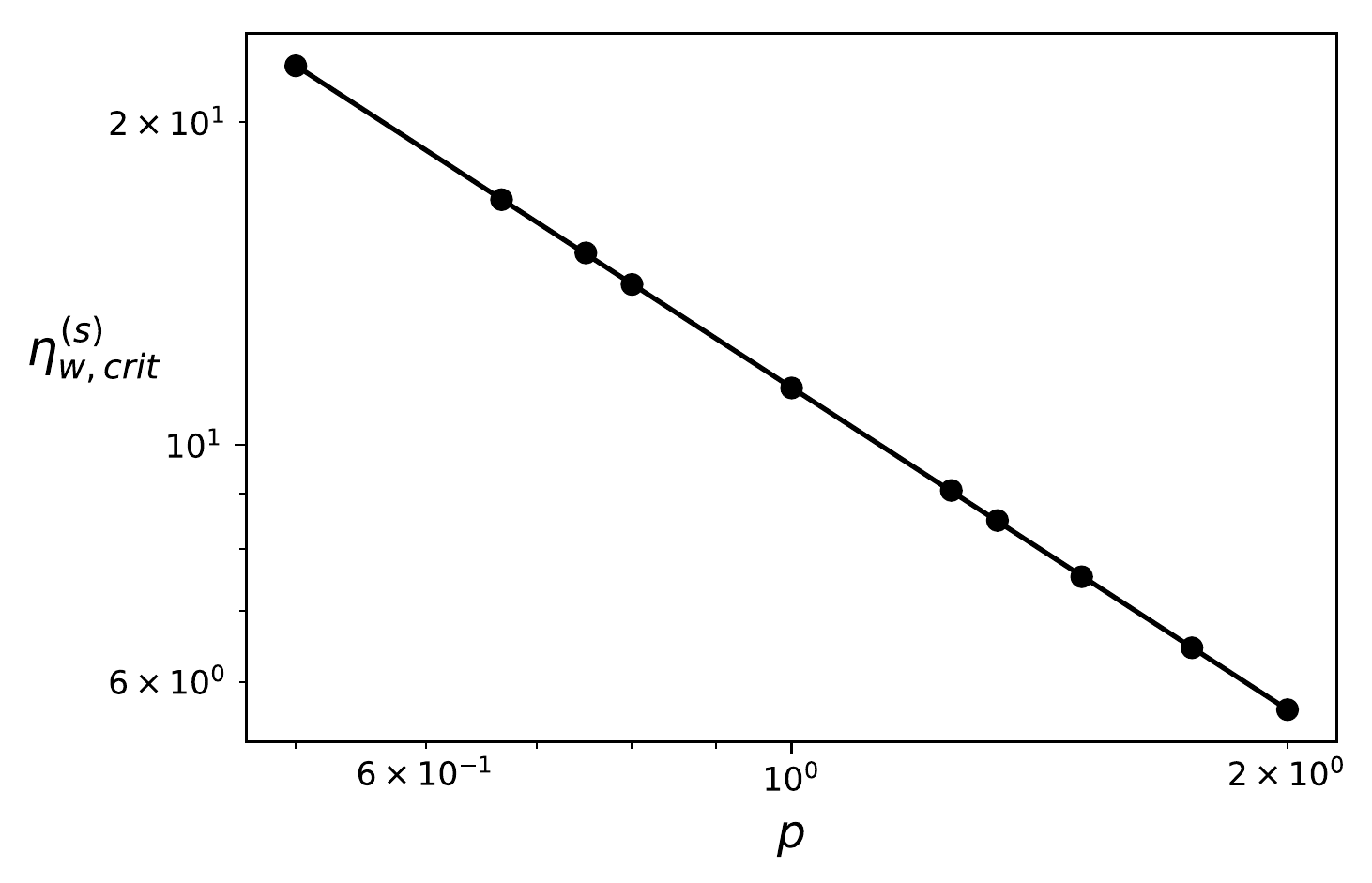}
\caption{The critical wind efficiency $\eta_{w,crit}$ as a function of the power-law index, $p$, in accordance with eq. \ref{eq7}.}
\label{fig:sl}
\end{figure}

Table ~\ref{tab:1} summarizes the combination of $p$ and $C^\prime$ (for $\Gamma=0$) values used in our simulations. The wind to accretion ratio sets the stability of the disk. The efficiency factor $\eta_w$ is a function of time and hence, we consider the value of $\eta_w$ at the beginning of the perturbation. The value of $\eta_w$ decreases with increasing $p$ with a slope of about -1 on a log-log plot (Fig. \ref{fig:sl}). This confirms the inverse relationship between $\eta_{w,\textrm{crit}}$ and $p$, that we discussed in $\S$\ref{sec:2.a} and demonstrates that the disk is easily destabilized if the wind is more strongly coupled to accretion (i.e. higher $p$).

\begin{table}[ht]
\caption{Summary of all parameters ($C^\prime$ and p) used in simulations for the model described by eqn \ref{eq7}. The subscripts d, crit and g denote values of $C^\prime$ for which the oscillations decay, persist and grow, respectively. The value of $C_{crit}^\prime$ increases with increasing $p$. In Fig. \ref{fig:sl}, we use the corresponding $C^\prime_{crit}$ to plot the wind efficiency $\eta_{w,crit}$, which decreases with increasing $p$.}
\begin{center}
\begin{tabular}{c |  c |   c |   c}
\hline
p&  $C^\prime_{d}$ & $C^\prime_{crit}$ & $C^\prime_{g}$  \\
\hline
0.5 & 4.5 & 4.95 & 5 \\
2/3 & 6 & 7.05 & 7.5 \\
3/4 & 8 & 8.6 & 9 \\
4/5 & 9 & 9.5 & 9.9 \\
1 & 11 & 11.3 & 14.4 \\
5/4 & 18.5 & 20 & 24 \\
4/3 & 21 & 22.8 & 24 \\
3/2 & 20.5 & 26 & 30.5 \\
7/4 & 37 & 39 & 40.5 \\
2 & 30.5 & 48.5 & 50.5 \\\hline
\end{tabular}
\end{center}
\label{tab:1}
\end{table}

Table \ref{tab:2} contains our parameter survey for the variable $R_L$ model. The $\Gamma$ factor strongly controls and alters the outcome of disk evolution. In particular, we identified new cases for high $\Gamma$ ($\Gamma\geq 0.7$). In low $\Gamma$ cases ($\Gamma\sim 0.2$), we observe some deviation from the classical cases of stable oscillations.

\begin{figure}[htb!]
\includegraphics[scale=0.55]{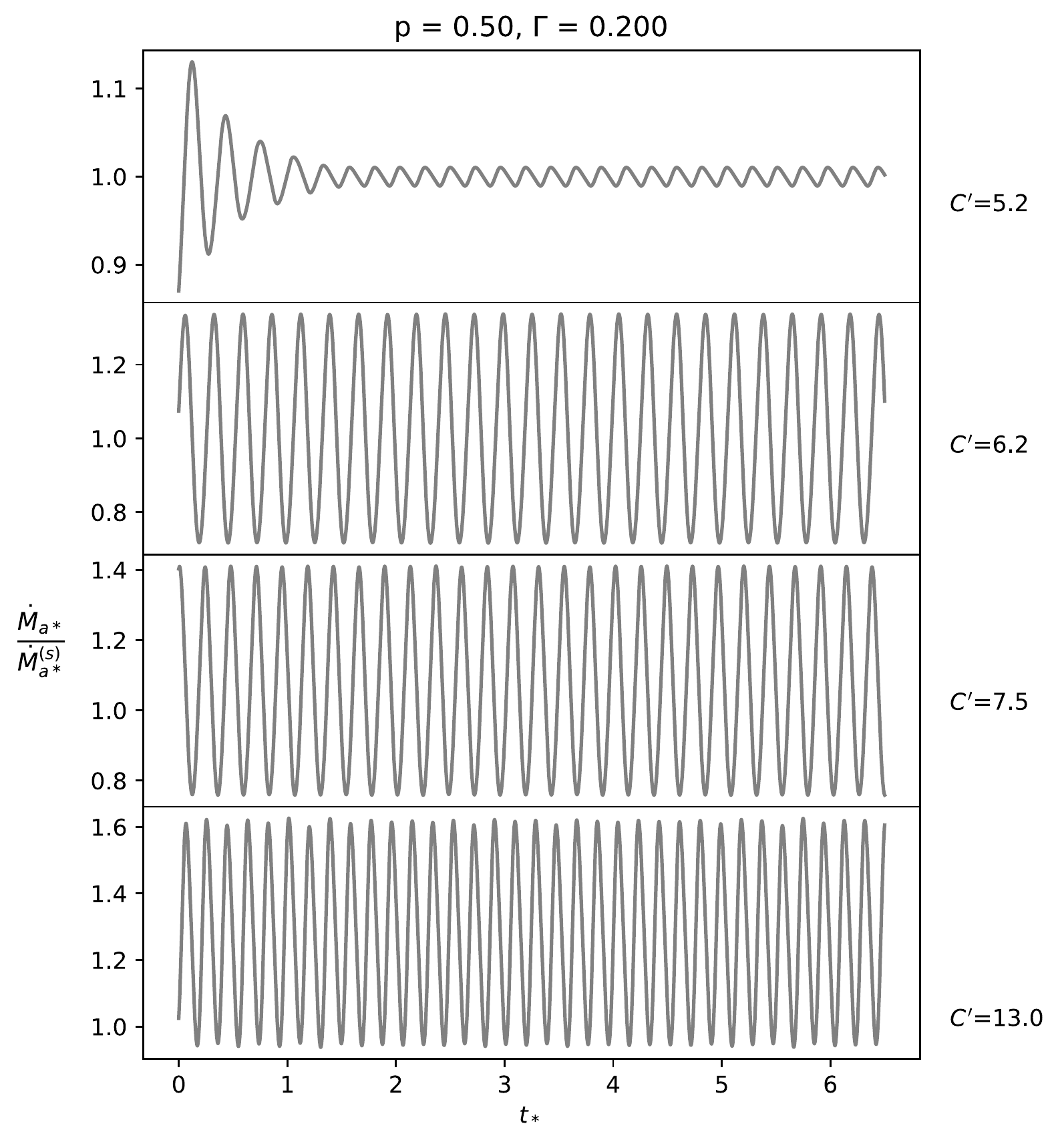}
\caption{Examples of the time evolution of the accretion rate for $\Gamma=0.2, p=0.5$,
and various $C^\prime$. The top panel shows an example of an initial decay that is followed by a constant amplitude oscillation. The second, third and fourth panels depict the range of $C_{crit}^\prime$ that lead to stable oscillations.}
\label{fig:g2}
\end{figure}

We limit our presentation to 3 cases, $p = 0.5, 1$ and $2$. The case $p = 1$ corresponds to the special case studied in \citetalias{Shields}. The other two cases are representative of the lowest and highest $p$ values considered (see Table \ref{tab:1}). For $0.001\leq \Gamma \leq 0.1$, the disk behaviour does not deviate much from the classical behaviour depicted in Fig. \ref{fig:shields result}. However, for $\Gamma\geq0.2$, we find some new results. We describe these cases in detail below.

\begin{figure}[htb!]
\includegraphics[scale=0.55]{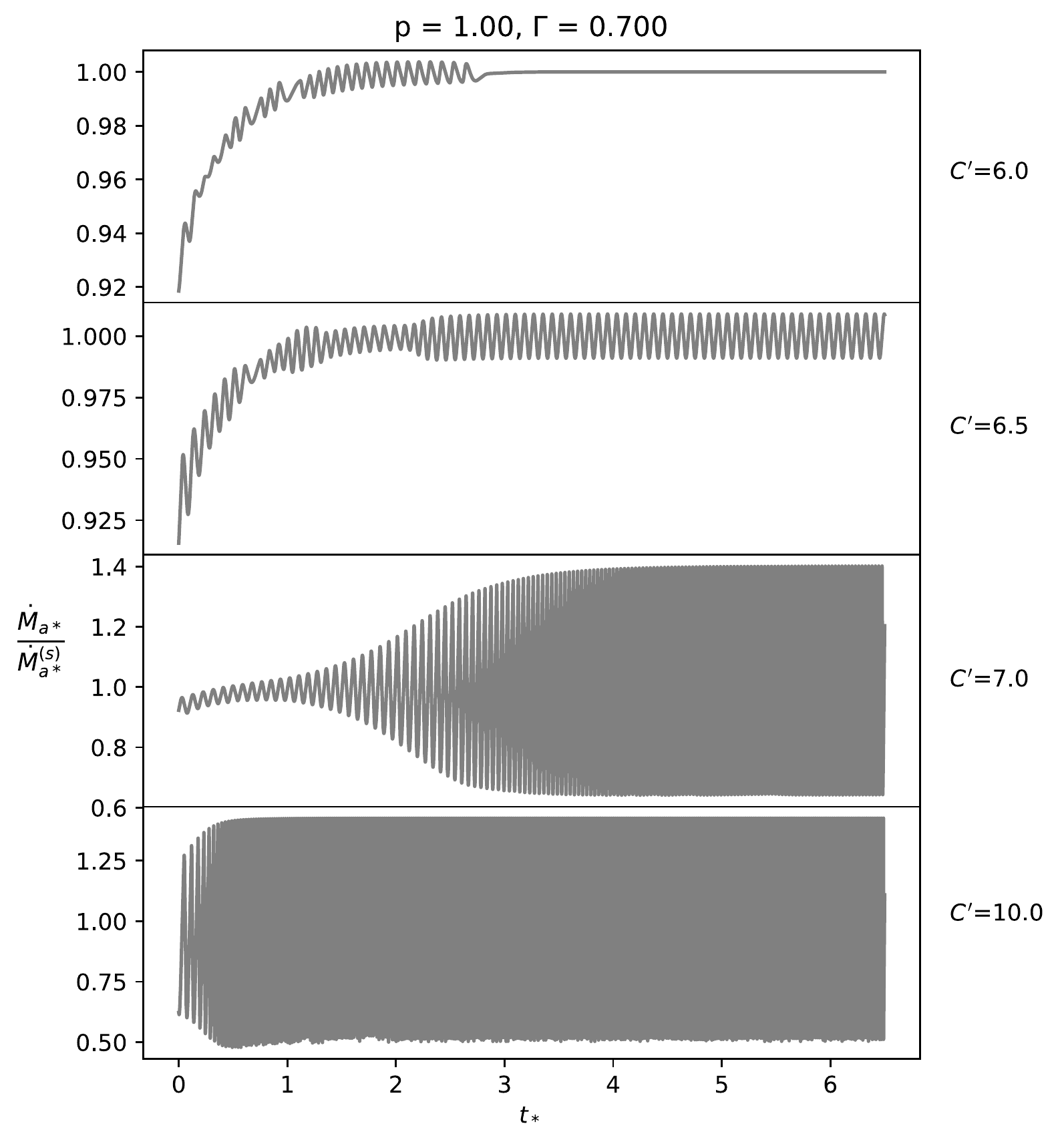}
\caption{As Fig. \ref{fig:g2} but for $\Gamma=0.7, p=1$. The top three panels indicate a damped high amplitude oscillation superposed on the fundamental mode oscillating around the steady state value of 1. The lowest panel shows growing oscillations that saturate but has a very high frequency of oscillation, which increases with time.}
\label{fig:g4}
\end{figure}

\begin{itemize}
    \item $\Gamma$=0.2 and 0.3: We observe more than one $C^\prime_{crit}$ value resulting in stable oscillations. For example, constant amplitude oscillations occur when $C_{crit}^\prime$ is between 6.2 and 12. The lowermost panel of Fig.~\ref{fig:g2} shows how the sinusoidal nature of the stable oscillations starts to change for $C^\prime\gtrsim 13$. We also find that for increasing $C^\prime$, the frequency of oscillation starts to increase. For $\Gamma=0.3$, the results are similar to those of $\Gamma=0.2$.
    \item $\Gamma=0.4$: For $p=0.5$ and $p=1$, we find a single value of $C^\prime_{crit}$. For $p=2$, we see that $C_{crit}^\prime$ lies between 130 and 300. The range of $C_{crit}^\prime$ and the critical value of $C^\prime$ rise considerably. 
    \item $\Gamma = 0.5$: For all $p$ values, we obtain single-valued $C^\prime_{crit}$. We note that $C^\prime_{crit}$ for $\Gamma=0.5$ is less than that for $\Gamma=0.4$. This result holds true for all three p-values investigated.
    \item $\Gamma = 0.7$: Fig. \ref{fig:g4} summarizes the distinct cases obtained for $p=1$. We see that up to a certain value of $C^\prime$, the accretion rate initially increases and at the same time oscillates with relatively high frequency and small amplitude. In the case of $C^\prime=6.5$, we see that eventually the fundamental oscillation dominates and steadies around 1. As $C^\prime$ increases, we see a growing oscillation that eventually saturates.
    \item \textbf{$\Gamma = 0.9$:} The behaviour is similar to the $\Gamma=0.7$ case. For higher $C^\prime$, the frequency increases rapidly with time.
\end{itemize}

The amplitude of oscillations in $\dot{M}_{a*}$ remain constrained to $\sim 50\%$ of its steady state value in all our simulations. This can be mostly attributed to the amount of matter available to be launched as wind and also the constraint on the lowest possible launching radius. In all of the cases studied, the amplitude of oscillation increases for increasing $\eta_w$.

As stated in $\S$\ref{sec:2.b}, we have chosen a spatial resolution of $N_x = 200$ for the above cases. Our resolution study showed that there are quantitative changes in most cases as well as qualitative changes in some of the extreme cases. For example, for $p=1, C^\prime=6, \Gamma=0.7$, the oscillations cease earlier for $N_x=400$, while for $N_x=800$, they persist for a longer number of time steps. In addition, the amplitude of oscillations also decrease with increasing resolution. For higher $N_x$, there is a clear tendency towards convergence. This resolution study was conducted for several other cases in our parameter survey. Convergence was evident for most cases with $\Gamma\leq 0.7$. For higher $\Gamma$, the behaviour was more erratic and unpredictable for different $N_x$. We restricted our $N_x$ to 200 despite this fact, since the nature of oscillations and the feedback on $\dot{M}_w$ remained unaffected from a qualitative perspective.

\begin{figure}[ht]
\includegraphics[scale=0.55]{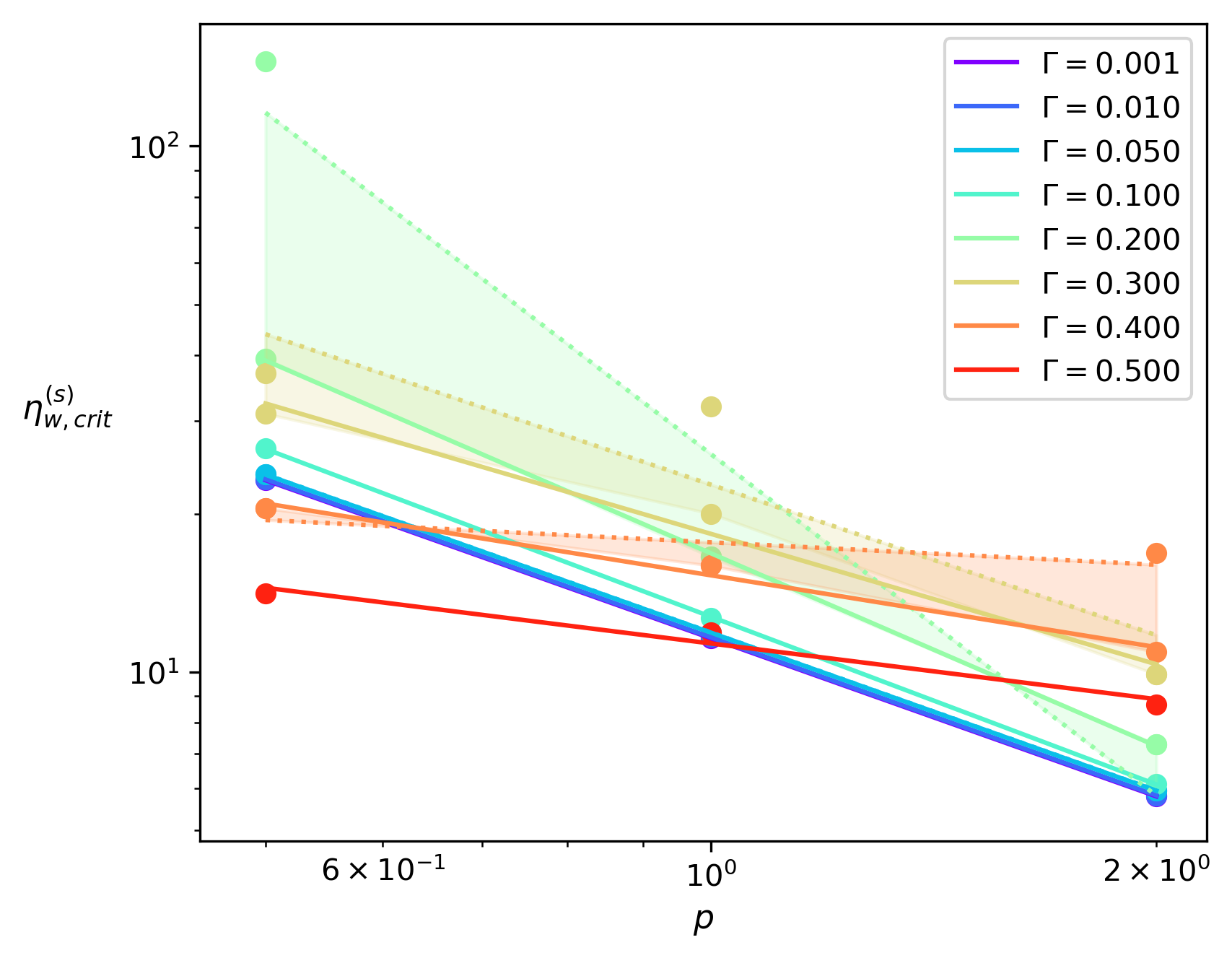}
\caption{The steady state critical wind efficiency $\eta_{w}$ vs $p$ for different $\Gamma$ values. For some $\Gamma$, stable oscillations occur not just for a single value of $\eta_w$, but for a range of $\eta_w$ (see Fig. \ref{fig:g2} for some examples, e.g. second, third and fourth panels there). We shaded the regions for $\Gamma$s where this happens.}
\label{fig:g1}
\end{figure}

In Fig. \ref{fig:g1}, we show the steady state critical $\eta_w$-$p$ relation for various $\Gamma$. We find that the slope of this relation is nearly constant for $\Gamma\leq 0.2$. For higher $\Gamma$, the slope changes and more than one $\eta_{w,crit}$ exists. For such cases we have shaded the region between all the possible straight line fits. For $\Gamma>0.5$, oscillations are distinctly different from that of the classical cases and we cannot group them under simple categories (see above and Fig. \ref{fig:g4}). We do not plot these points in Fig. \ref{fig:g1} or list them in our table of classification of disk oscillations. 

\begin{figure}[ht]
\includegraphics[scale=0.55]{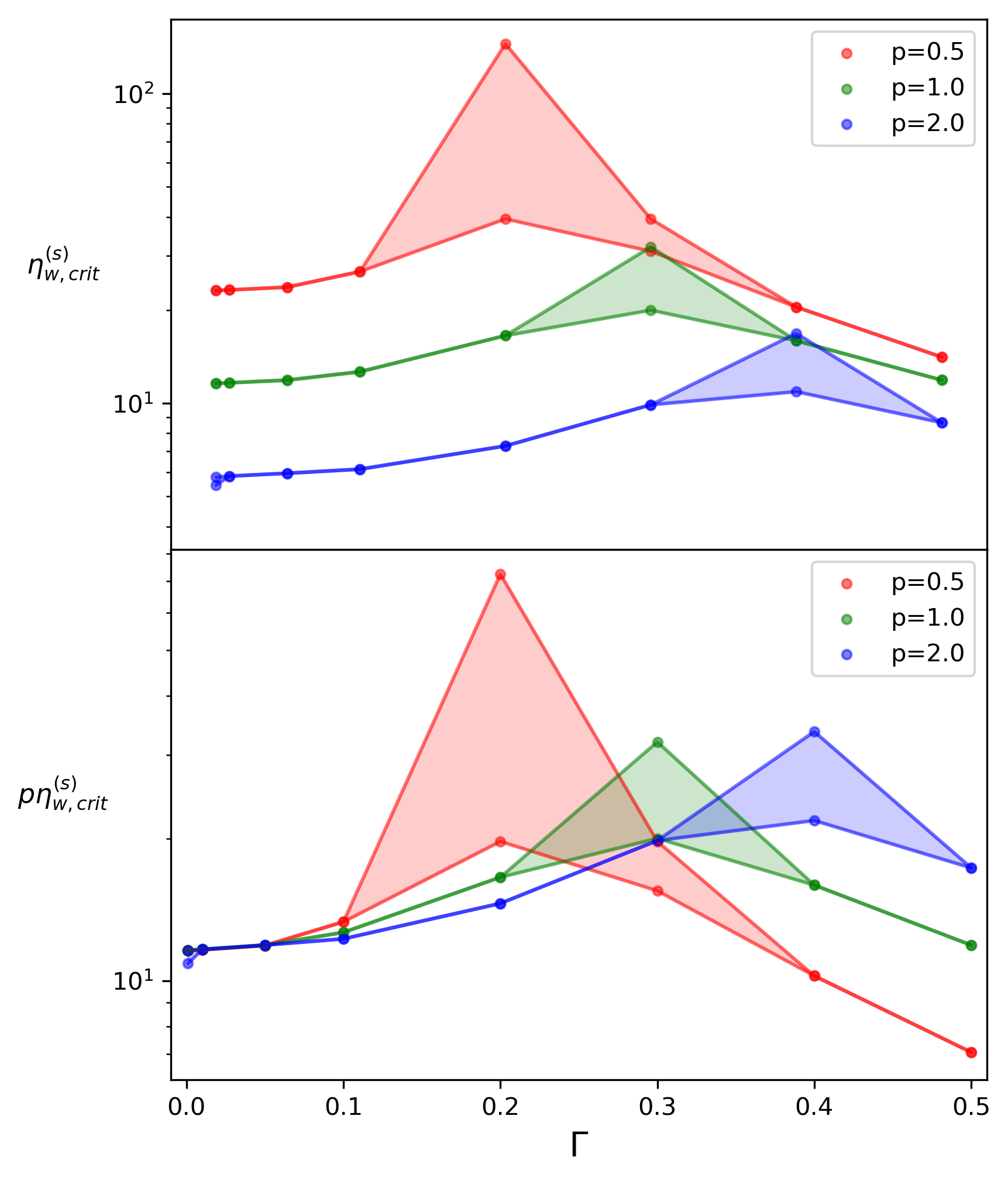}
\caption{The steady state critical wind efficiency $\eta_{w,crit}$ vs $\Gamma$ for different $p$ values (top panel). In the bottom panel, we plot $p\eta_{w,crit}$ vs $\Gamma$ for different $p$ values. Once again, the shaded region highlights the possible $\eta_{w,crit}$ values for a particular combination of $p$ and $\Gamma$.}
\label{fig:g3}
\end{figure}

To visualize our results for the wind efficiency in a different way, we also plot steady state critical $\eta_w$ as a function of $\Gamma$ for different $p$ values (the upper panel of Fig. \ref{fig:g3}). The curves for different $p$ values generally resemble each other. For $\Gamma\leq 0.2$, $\eta_{w,crit}$ increases with increasing $\Gamma$. For higher $\Gamma$, $\eta_w$ decreases with increasing $\Gamma$, in all three $p$ cases. To more directly compare the results for various $p$, in the bottom panel of Fig. \ref{fig:g3}, we plot $p\eta_{w,crit}$ vs $\Gamma$. For $\Gamma<0.1$, we see that $p\eta_{w,crit}$ is nearly constant, which is what we concluded from Fig.~\ref{fig:sl}.

\section{Concluding Remarks}
\label{sec:4}

The study of disk winds in the context of state transition has evoked a lot of interest over the past decades \citep[e.g.][]{RF05, RF06}. The wind launching mechanism and location may lead to time variability, the effects of which could be coupled to the signatures of mass accretion by the accretor. We explore this aspect through a self-regulated accretion disk. We found that our model is unlikely to explain state transitions in accretion disk spectra. However, it could be responsible for persistent small amplitude regular single-mode oscillations in the mass accretion rate. Admittedly, our treatment is very simple. We use one viscosity law in all our analyses, $\nu_* = R_* = x^2$. We did not consider thermodynamic effects. We also did not incorporate magnetic fields in our analysis. Moreover, we assumed a $\delta$-function model for the wind, which exaggerates the role of mass removal.

Despite the simplifications, this analysis might be a stepping stone towards developing models that can account for the existence of state transition signatures in accretion disks. Time variability in accretion rates may be explained by physical processes considered here. They are rich in features and may hold the key to understanding the coupling  forces operating in an accretion disk. For example, the study of GRS 1915+105 by \cite{N11} is a very detailed analysis in that direction. They nicknamed the oscillations in X-ray spectra as the `heartbeat' state and conducted a study of the geometry of the accretion disk using the X-ray continuum and emission lines in the optical spectrum. They demonstrate a strong correlation between mass loss in the form of wind and oscillations in the accretion rate that would explain the long-term effects in the disk.

A study of the same source, GRS 1915+105, by \citet{Z16} considered in detail the reflection spectrum during the oscillatory phase of the source. Their calculations indicate winds being launched from very small disk radius which remains unchanged during this phase. Our variable $R_L$ model allows the launching of wind from as close as the innermost disk region for high $\Gamma$ cases. The time evolution of the accretion rate from our simulation does not indicate any distinct spike or sharp flare, which leads us to conclude that the system does not produce outbursts. \cite{HP17} used their photoionization modelling of the SEDs to argue that thermally driven winds may hold the key to explaining the state changes in such systems. On the other hand, \citet{N13} goes on to demonstrate how heavy outflows not only quench the disk, thus affecting the formation of jets and causing state transition, but also influence the further production of winds. Disk variability and the state changes could also be caused or affected by instabilities of the disk itself. For instance, \cite{J02} studies the effect of radiation instabilities leading to limit-cycle behaviour and modulations in accretion rates. As opposed to our model, this radiation-driven instability may result in sharp spikes indicating a high outflow from the system.

There are a number of studies related to fast outflows with high mass loss rates leading to state transitions and subsequent detection of jets in the system \citep{NL09, K13, G19}. A similar situation has been discussed in \citet{RF19}, where a particular case of V404 Cyg indicates the presence of massive outflows, almost 2 orders of magnitude higher than central mass accretion rate $\dot{M}_a$. They speculated that these outflows are produced by radiation-driven winds coupled with classical thermal winds. This work stands out as providing direct observational evidence of powerful outflows leading to a quenching of accretion. The optical $H_\alpha$ line profile clearly indicates disk contraction following the massive outflow phase, consistent with what we would expect happens when irradiation is reduced. Our analysis of the $\Sigma$ radial profile showed that $\Sigma$ approaching 0 is responsible for saturating the oscillations.

Recently, \citet{T19} showed that the detection of a blue-shifted line from a black hole binary source H1743-322 strongly suggests a thermal-radiative disk wind. Their work indicates a disappearing wind in the hard state which could be attributed to the shadowing of outer disk region by the inner corona. They also went on to state that the absorption features in other black hole sources such as GRS 1915+105 and GRO J1655-40 are most likely due to thermal-radiative winds as opposed to previously speculated magnetic effects. Another recent paper \citet{D19} studies the effect of thermal-viscous instability on the light curves and stability diagrams associated with black hole binary systems. Additionally, they consider a fraction of the X-ray irradiation to be scattered by the wind and partially impinge on the outer disk regions. This has a stabilizing effect and can explain the shortened outbursts but cannot explain the rapid decay of outbursts. They studied a particular BHXB, GRO J1655-40, for which their model was able to reproduce the observed features of the light curve. They speculate that magnetic fields would need to be considered for a more promising explanation for the outbursts.

These studies indicate several possible aspects of our simplistic approach towards the study of wind-accretion coupling. The criteria for instability derived by \citetalias{Shields} is often invoked when discussing consequences of observed or model disk winds \citep[e.g.,][]{2010Lu}. The main conclusion of our work is that upon satisfying this criteria, a disk wind might not be responsible for large scale variations in luminosity because instability saturates at a relatively low level in terms of $\dot{M}_a$. However, it could result in a harder-to-detect significant reduction of $\Sigma$ at large radii.  

\acknowledgements

This work  was supported by NASA under ATP grant 80NSSC18K1011. 
We thank  Drs. Tim Waters and Rebecca Martin for their valuable comments and discussions.

%\bibliography{main}

\begin{table*}[h]
\begin{center}
\begin{tabular}{| c | >{\centering}m{2cm} | >{\centering}m{2.5cm} |  >{\centering}m{2.5cm} |  >{\centering}m{2.5cm} | c|}%C{2.5cm} |}
\hline
\multicolumn{6}{|c|}{\textbf{p=0.5}} \\ \hline
\textbf{$\Gamma$}& \textbf{$C^\prime_{sd}$} & \textbf{$C^\prime_d$} & \textbf{$C^\prime_{crit}$} & \textbf{$C^\prime_g$} & \textbf{$C^\prime_{sg}$} \\
\hline
0.001 & - & 3.5 $\longrightarrow$ 4.7$^*$ & 4.71 & 4.715 $\longrightarrow$ 4.75 & 4.8 \\
0.01 & - & 4 $\longrightarrow$ 4.71 & 4.72 & 4.8 $\longrightarrow$ 5 & - \\
0.05 & - & 4.7 $\longrightarrow$ 4.75 & 4.772 & 4.8 $\longrightarrow$ 5 & - \\
0.1 & - & 5 $\longrightarrow$ 5.05 & 5.067 & 5.1 $\longrightarrow$ 5.2 & - \\
0.2 & 5.2 $\longrightarrow$ 5.5 & 5.8 $\longrightarrow$ 6 & 6.2 $\longrightarrow$ 12 & - & \\
0.3 & - & 4 $\longrightarrow$ 5.45 & 5.48 $\longrightarrow$ 6.2 & 6.5 $\longrightarrow$ 9 & 10 $\longrightarrow$ 15 \\
0.4 & 4 $\longrightarrow$ 4.417 & - & 4.418 & - & 4.419 $\longrightarrow$ 5 \\
0.5 & 3.4 $\longrightarrow$ 3.6 & -$^{**}$ & 3.63 & - & 3.7 $\longrightarrow$ 4 \\
\hline
\multicolumn{6}{|c|}{\textbf{p=1}} \\ \hline
\textbf{$\Gamma$} & \textbf{$C^\prime_{sd}$} & \textbf{$C^\prime_d$} & \textbf{$C^\prime_{crit}$} & \textbf{$C^\prime_g$} & \textbf{$C^\prime_{sg}$} \\
\hline
0.001 & - & 11.4 & 11.6 & - & 11.65  \\
0.01 & - & 11.6 & 11.65 & 11.7 & - \\
0.05 & - & 11.8 & 11.87 $\longrightarrow$ 11.9 & 11.95  & - \\
0.1 & - & 12.6 & 12.65 & 12.7 & - \\
0.2 & - & 16.5 & 16.55 & 16.6 & - \\
0.3 & 18 &  - & 20 $\longrightarrow$ 32 & - & 45 \\
0.4 & 15 &  - & 15.93 & - & 20 $\longrightarrow$ 25  \\
0.5 & - & 11.8 $\longrightarrow$ 11.88 & 11.89 & - & 11.9 $\longrightarrow$ 11.95  \\
\hline
\multicolumn{6}{|c|}{\textbf{p=2}} \\ \hline
\textbf{$\Gamma$}&  \textbf{$C^\prime_{sd}$} & \textbf{$C^\prime_d$} & \textbf{$C^\prime_{crit}$} & \textbf{$C^\prime_g$} & \textbf{$C^\prime_{sg}$} \\
\hline
0.001 & - & 39.4 & 35 $\longrightarrow$ 39.3 & 39.5 $\longrightarrow$ 40 & 50  \\
0.01 & - & 11 $\longrightarrow$ 39.5 & 39.7 & 40 & - \\
0.05 & - & 39.7 $\longrightarrow$ 41 & 41.3 & 41.5 $\longrightarrow$ 42  & - \\
0.1 & - & 42 $\longrightarrow$ 43.5 & 43.7 & 44 & - \\
0.2 & - & 43 $\longrightarrow$ 55 & 60.35 & 62 $\longrightarrow$ 65 & - \\
0.3 & - & 60 $\longrightarrow$ 100 & 107.8 & 110 $\longrightarrow$ 120  & - \\
0.4 & - & 110 & 130 $\longrightarrow$ 300 & 500 & - \\
0.5 & 78 & 80 $\longrightarrow$ 83 & 83.7 & 85 & 100  \\
\hline
\end{tabular}
\caption{Summary of parameter survey. $C_d^\prime, C_{crit}^\prime, C_g^\prime$ hold the same meaning as described in the Table~\ref{tab:1}. The two new types of outcome, $C_{sd}^\prime$ and $C_{sg}^\prime$, denote oscillations that decay/grow for a few time scales before stabilizing and oscillating with a constant amplitude. In Fig. \ref{fig:g1} and \ref{fig:g3}, we use the corresponding $C^\prime_{crit}$ to plot the wind efficiency $\eta_{w,crit}$.}
\end{center}
\label{tab:2}
\footnotesize{$^*$The $\longrightarrow$ denotes the range of tested values that fall under a certain category. The limits of the range are not absolute but give a more or less general idea of the behaviour within those $C^\prime$ values.\\
$^{**}$ Empty cells denote parameter combination for which we do not find the outcome in question. This does not rule out the possibility of such a case or of any other new cases. $C_{crit}^\prime$ is the only value that was important for our analysis and the parameter survey was focused on finding the condition for stable (or critical) oscillations.}
\end{table*}
\end{document}